\documentclass[prl,preprint,showpacs,aps]{revtex4}

\usepackage{graphicx}
\usepackage{dcolumn}
\usepackage{bm}

\begin{document}

\begin{titlepage}

\title{Modulating transmission properties of nanoscale transistors
by dipole effects near contacts}

\author{Li Yang, Jian Wu\footnote{Author to whom any correspondence
should be addressed. Email address: wu@castu.tsinghua.edu.cn},
Wenhui Duan, and Bing-Lin Gu}
\address{Department of Physics and Center for Advanced Study,
Tsinghua University, Beijing 100084, People's Republic of China}

\date{\today}

\begin{abstract}

We theoretically demonstrate that a dipole layer on the electrode
can modulate the transmission properties of nanoscale devices by
influencing the contact properties, through first principles
simulations on carbon nanotube based field effect transistors. The
dipole layer is realized by potassium adsorption on Au electrodes,
which shifts the electrostatic potential at the near contact region
significantly. The dipoles parallel to the direction of the bias
voltage may act as a supplement to the effect of gate voltages,
while the perpendicular dipoles may modify the interface barrier of
the contacts.

\end{abstract}


%85.35.-p Nanoelectronic devices
%73.40.Cg: Contact resistance, contact potential
%73.22.-f Electronic structure of nanoscale materials: clusters, nanoparticles, nanotubes, and nanocrystals
%85.35.Kt Nanotube devices

\maketitle

\draft

%\vspace{2mm}

\end{titlepage}

The miniaturization of devices is an irresistible trend in
industrial manufacture and becomes a hot research topic. Nanoscale
devices, such as field-effect transistors (FETs) based on nanotubes
\cite{tans,martel} or nanowires \cite{cui}, have attracted more and
more interests. In general, the transmission properties of FETs are
controlled through a gate voltage. However, with the device size
decreasing to nanoscale, the potential profile near the electrodes
may play a central role in electron transport behavior, especially
for systems with metal-semiconductor contacts. In such systems, the
potential pinning effect of the metallic electrodes will lead to a
change of the potential profile near the contacts as shown in Fig.
1(a) \cite{heinze,wind} under the gate voltage, which is not
favorable for the quantum tunneling of electrons, and thus make the
conventional modulation by the gate voltage less effective. Though
it has been demonstrated that the device performance may be improved
by using needle-like contacts \cite{heinze} or using small gate
segments near the electrodes \cite{wind} or adopting special gate
configuration \cite{apl5038}, the problem is only partially solved
since the potential pinning effect of the metallic electrodes can
not be fully overcome but only limited to a slightly smaller region.
Considering the fact that the potential pinning effect comes from
the metallic surface, modulating the property of the metallic
surface directly might be a better approach to improve the
performance of nanoscale devices.

\begin{figure}[tbp]%fig1
\includegraphics[width=7.0cm]{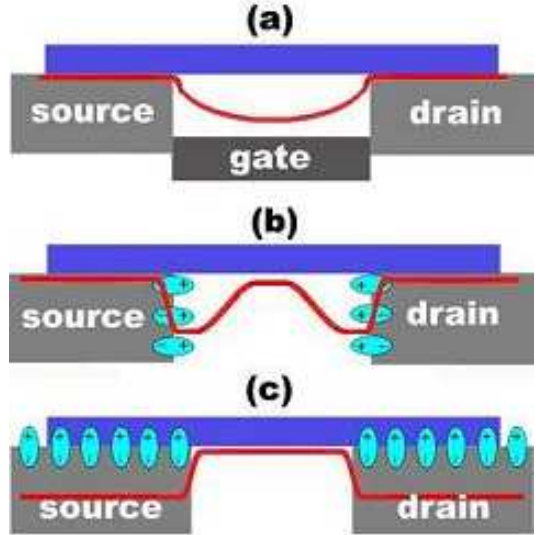}
\caption{(Color online) Sketch of the electrostatic potential (the
solid lines) on the sample induced by (a) a gate voltage, (b)
``parallel'' dipoles, and (c) ``perpendicular'' dipoles.
}\label{Fig1}
\end{figure}

In this letter, we demonstrate that the modulation of the
transmission properties of nanoscale transistors can be realized by
introducing a dipole layer on the electrodes. As schematically shown
in Fig. 1(b), an additional electrostatic potential induced by a
dipole layer on the lateral side of electrodes may significantly
change the electrostatic potential of the functional body near the
interface, and acts as a supplement to the effect of gates.
%Such additional electrostatic potential will greatly affect
%the contact properties.
With the potential pinning effect of the metallic surface balanced
or strengthened by the dipole layer, a significant modulation effect
can be expected in the system. Besides, the dipole layer on the top
side of electrodes will modify the electrostatic potential of the
functional body at the interface region as shown in Fig. 1(c) and
thus modify the interface barrier.

In general, the dipoles parallel to and perpendicular to the
direction of the bias voltage have very different effects on the
system. The ``parallel'' dipoles may act as a supplement to the
effect of gates; while the ``perpendicular'' dipoles may only modify
the interface barrier of the contacts. As a case study, we will
investigate the effect of the dipole layer, which is induced by
chemical adsorption on electrodes, on the contact properties of
carbon nanotube based field effect transistors (CNTFETs). It is
known that adsorption of alkali atoms on metal surface will lead to
electric dipole moments \cite{bonzel, ebinger, rangelov}. Here, we
will focus on the case of K adsorption on Au electrodes in CNTFET,
which has also been extensively studied in recent experiments
\cite{heinze, derycke, radosavljevic}. Specifically, we use the
lateral adsorption and top adsorption (as shown in Fig. 2) to
introduce ``parallel'' dipoles and ``perpendicular'' dipoles
respectively.

Our study is performed by numerical simulations using the plane
wave basis VASP code \cite{Vasp,Van} within the framework of
density functional theory (DFT) under local density approximation
(LDA). The planewave cutoff energy is 287 eV. Integration over the
Brillouin zone is done using $\Gamma$ centered Monkhorst-Pack
scheme \cite{monkhorst} with 1$\times$1$\times$1 and
2$\times$8$\times$1 for lateral and top adsorption respectively.
Relaxations of atomic positions were carried out until the forces
on each atom are less than 0.01 eV/\AA.

The K-adsorbed Au/CNT contact system studied is represented by the
supercell as shown in Figs. 2(b) (lateral adsorption) and 2(d) (top
adsorption) with periodic boundary condition, where a large enough
vacuum layer and dipole corrections \cite{neugebauer} in the z
direction are introduced to avoid non-physical interactions between
repeated images of the unit cell. (8,0) carbon nanotube and Au(100)
slab are used in our simulations. Test calculations are done with
both five and three Au (100) layers and no essential difference is
found in the properties concerned. Therefore, we take the slab model
of three Au(100) layers in the following discussion for both lateral
and top adsorption cases.

\begin{figure}[tbp]%fig2
\includegraphics[width=8.5cm]{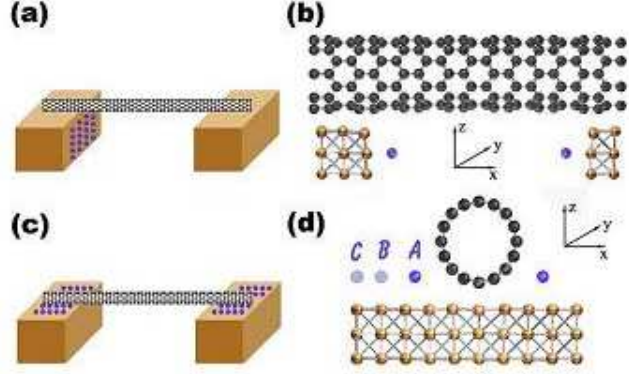}
\caption{Schematic illustration (a) and atomic geometry (b) of
Au(100)/CNT(8,0) contact with K lateral adsorption. Schematic
illustration (c) and atomic geometry (d) of Au(100)/CNT(8,0) contact
with K top adsorption. The gold, purple, and black balls represent
Au, K, and C atoms, respectively.
%One unit cell is illustrated in (b) and two
%unit cells are illustrated in (d).
}\label{Fig2}
\end{figure}

In most recent FET experiments, annealing is done before the K
adsorption process to obtain intimate contacts between electrodes
and the nanotube \cite{Auvray, yaish}. So, as the first step of our
simulation, a good contact between Au electrodes and CNT is obtained
by a structural optimization with the bottom layer of the Au slab
fixed and all other Au atoms and C atoms fully relaxed. By comparing
the binding energy for typical contact geometries such as DT (direct
on top), BM (bridge middle), and HC (hexagonal center), we obtain
the most stable geometry (BM geometry) of Au/CNT contact. Here the
lattice parameters of Au(100) slab along the tube axial direction
are elongated slightly to match that of (8,0) CNT while those in the
other two directions are equally compressed to make the volume a
constant as done in Ref. \cite{zavodinskya}.

Then we determine the contact structures associated with K
adsorption. It is found that K atoms can be chemically adsorbed on
clean Au(100) surface, with adsorption energy of up to 2.86 eV (at
the hollow sites), which is much larger than that on CNT (typically,
1.43 eV). This indicates that K atoms prefer to be adsorbed on the
electrodes rather than on the CNT during K doping treatment in
CNTFET experiments. For obtaining the favorable adsorption site of K
atom on Au surface in the Au/CNTFET system, the K atom is initially
put on different hollow sites with different distance from the CNT
and then the structures are optimized. The adsorption energies
obtained are 3.04 eV, 2.81 eV and 2.79 eV, respectively, for the
three adsorption sites A, B and C (as shown in Fig. 2(d)). This
means that the K atom prefers to stay at A site closest to the CNT.
In the optimized structure, the K atom also deviates from the center
of the hollow site and moves towards to the CNT. Therefore, in the
following simulations, we focus on the structures shown in Figs.
2(b) and 2(d), where the K atoms are placed around the hollow sites
of Au(100) surface near the (8,0) CNT, to demonstrate the dipole
effects.

\begin{figure}[tbp]%fig3
\includegraphics[width=8.5cm]{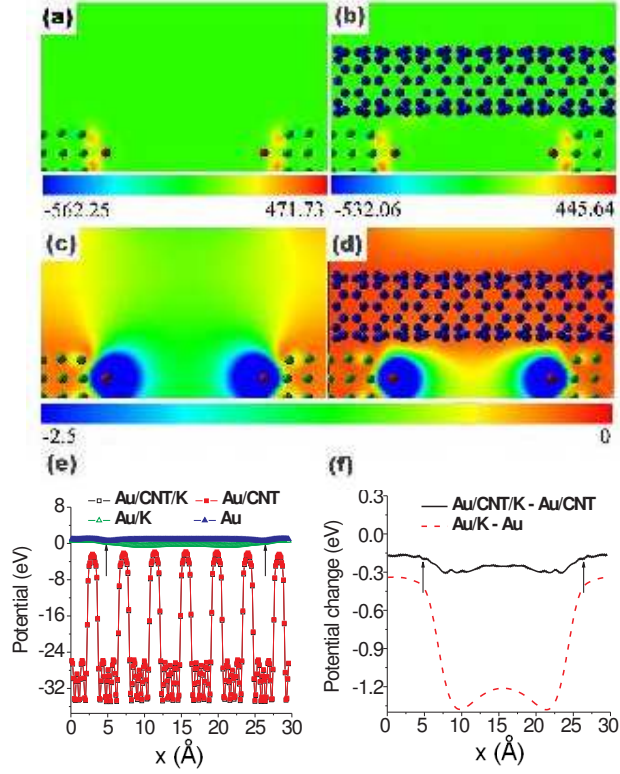}
\caption{(Color online) The physical property changes induced by K
lateral adsorption. (a) Charge difference, and (c) Local potential
difference between clean and K adsorbed Au surfaces. (b) Charge
difference, and (d) Local potential difference between clean and K
adsorbed Au/CNT contacts. (All the slices are selected passing
through the K atoms.) (e) Local potential along the line
perpendicular to the lateral surface and passing through the bottom
carbon atoms of CNT for the four systems: clean Au surface, K
adsorbed Au surface, pure Au/CNT contact, and K adsorbed Au/CNT
contact. (f) Local potential change after K adsorption along the
same line, for Au surface as well as Au/CNT contact. The green, red,
and blue balls represent Au, K, and C atoms respectively. The solid
arrows denote the edges of Au electrodes.}\label{Fig3}
\end{figure}

The lateral adsorption and top adsorption lead to ``parallel''
dipoles and ``perpendicular'' dipoles, respectively. First we
discuss the lateral adsorption case. To clearly show the effect of a
dipole layer on a device, we study the effect of a pure dipole layer
by taking away the CNT from the system shown in Fig. 2(b). Fig. 3(a)
shows the charge transfer in this case. It is obvious that almost
all the valence electrons of the K atoms transfer to the Au surface,
and this results in dipoles pointing from the surface to the K ions.
Note that the dipoles form a dipole layer due to the periodical
boundary condition used in the simulation, which is just what we
want. Such a dipole layer on electrodes will cause a downward shift
of the electrostatic potential nearby, as shown in Fig. 3(c). For a
better view, the local potential plotted along the line
perpendicular to the lateral surface and passing through the bottom
carbon atom sites of CNT is shown in Fig. 3(e) and the potential
change is shown in Fig. 3(f). It can be seen that the downward shift
of potential is up to 1.37 eV, a large value for most devices.

For a dipole layer of finite size (with characteristic length $L$),
the potential shift induced by the dipole layer at a distance $x$ is
nearly a constant when $x{\ll}L$, since the dipole layer can be
regarded to be infinitely extended in this case. While the potential
shift decreases as $x^{-2}$ when $x{\gg}L$, where the dipole layer
can be regarded as a point dipole. Furthermore, the characteristics
of the dipole layer can be modulated by changing its dipole strength
and the layer size. For adsorption induced dipoles in CNTFET
systems, this can be realized by changing the adsorption coverage
and electrodes size. Thus, the potential drop caused by the dipole
layer in the near interface region may be used to modulate the
transmission properties of a nanoscale device, as a supplement to
the effect of gates. Moreover, the change of the potential profile
(schematically shown in Figs. 1(a) and 1(b)) is now limited to a
length of approximately 5 \AA~ (see Fig. 3(f)), much smaller than
the typical length (about 5 nm) under a gate voltage about 10 V
\cite{heinze}. Therefore, such a supplement is quite effective.

Next, we quantitatively show the effect of the potential shift on
the CNT induced by the dipole layer. When a CNT is included in the
system, most of the valence electrons of K atoms transfer to the Au
surface while little to the CNT (as shown in Fig. 3(b)).
Consequently, the dipole layer formed between K ions and the Au
surface still plays the main role. This dipole layer will introduce
an additional electrostatic potential as discussed above, and, at
the same time, cause a charge redistribution between Au electrodes
and the CNT. This charge redistribution counteracts the effect of
the dipole layer and reduces the potential change on the CNT (as
shown in Fig. 3(d)). Fig. 3(e) shows the local potential on the CNT
along the line perpendicular to the lateral surface and passing
through the bottom carbon atoms of the CNT,  and Fig. 3(f) shows the
related potential change. In comparison with the case of the pure
dipole layer, the charge redistribution reduces the potential change
effect, but the net effect of the dipole layer on the CNT is still
found to be a downward potential shift up to 0.37 eV, indicating a
significant change of contact properties. It should be noted that
this effect will be more significant with larger coverage ratio of K
atoms. Furthermore, the potential shift on the CNT is somewhat
underestimated since the electrostatic response from other CNTs is
included in our simulations due to the periodical boundary
condition, while only one or very few CNTs are presented in most
real systems. Therefore, through our calculation, we demonstrate
that ``parallel'' dipoles on electrodes can effectively complement
the conventional effect of gate voltage; such complement can
significantly change the contact properties and thus modify the
transmission properties of CNTFETs.

\begin{figure}[htbp]%fig4
\includegraphics[width=9.0cm]{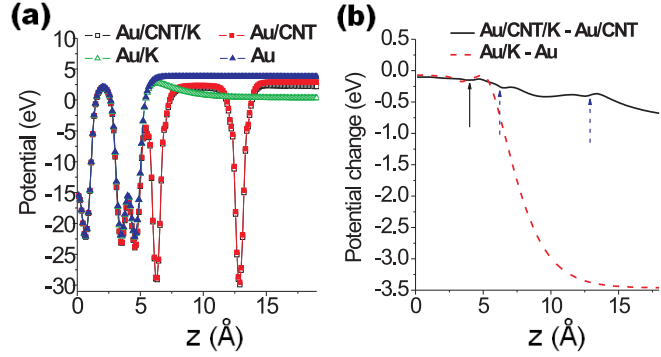}
\caption{ (Color online) The physical property changes induced by
K top adsorption. (a) Local potential along the line perpendicular
to the top surface and passing through the bottom carbon atoms of
CNT for the four systems: clean Au surface, K adsorbed Au top
surface, pure Au/CNT contact, and K adsorbed on top surface of
Au/CNT contact. (b) Local potential difference along the same
line, for Au surface and Au/CNT contact. The solid arrow denotes
the edge of Au electrode, and the dashed arrows denote the edges
of CNT.}
\end{figure}

Although both the top surface adsorption and the lateral adsorption
result in the dipole layer, they have quite different effects on the
device. Similar to the lateral adsorption case, the adsorption of K
atoms on Au top surface will result in a dipole layer formed by
dipoles pointing from the surface to the K ions, and thus cause a
downward shift of the electrostatic potential. Such additional
potential will lead to a charge redistribution between Au electrode
and the CNT, which counteracts the potential change and reduces the
potential change on the CNT.

Fig. 4 shows the local potential and its change induced by K
adsorption for Au surface and Au/CNT contact system with K top
adsorption. Here, the potential is plotted along the line
perpendicular to the top surface and passing through the bottom
carbon atom of CNT. It is found that the potential on the CNT is
greatly shifted: the potential of the bottom carbon atoms of CNT is
downward shifted up to 0.25 V and the potential of the topmost
carbon atoms of CNT is downward shifted up to 0.39 V. This
represents the magnitude of interface barrier (between the metal and
the CNT) modification, and might be underestimated for the same
reason as we discussed in the lateral adsorption case. Therefore,
``perpendicular'' dipoles may also contribute to the transmission
modulation of nanoscale transistors by modifying the interface
barrier. It should also be noted that due to the fact that effect of
the dipole layers is orientational dependent, it can only affect the
interface barrier perpendicular to the plane of the dipole layer.

In conclusion, we have demonstrated that introducing dipole layers
on the electrodes can significantly modulate the contact properties
of nanoscale devices. Specifically, via K adsorption on Au
electrodes, we introduced dipole layers parallel or perpendicular to
the direction of the bias voltage, and quantitatively revealed their
effects on the contact properties of CNTFETs. The ``parallel''
dipoles induce a change of the electrostatic potential of the
functional body (especially near the contact), as a supplement to
the effect of gates; while the ``perpendicular'' dipoles modify the
interface barrier of the contacts.
%and consequently modifies the potential profile of the CNT.
%K adsorption on electrodes forms a dipole layer which can
%significantly shift the vacuum level of the electrodes and
%thus influence the energy levelalignment of the interface.
Such striking effects of the dipole layer might play an important
role in optimizing contact properties in the design of
nanoelectronic devices.

This work was supported by National Natural Science Foundation of
China (Grant No. 10325415 and 10274038), the Ministry of Science and
Technology of China (Grant No. 2002AA311153), and the Ministry of
Education of China (Grant No. 200017 and NCET-04-0085).

\end{document}